\documentclass[conference]{IEEEtran}
\IEEEoverridecommandlockouts
\usepackage{cite}
\usepackage{array}
\usepackage{amsmath,amssymb,amsfonts}
\usepackage{graphicx}
\usepackage{textcomp}
\usepackage{xcolor}
\usepackage{algpseudocode}
\usepackage{algorithm}
\usepackage{verbatim}

\def\BibTeX{{\rm B\kern-.05em{\sc i\kern-.025em b}\kern-.08em
    T\kern-.1667em\lower.7ex\hbox{E}\kern-.125emX}}
\begin{document}
\renewcommand\arraystretch{1.3}
\title{An Algorithm for Transmitting VR Video Based on Adaptive Modulation\\
{}
\thanks{J. Feng, Y. Wu, G. Zhai, N. Liu, and W. Zhang are with the Department of Electronic Engineering, Shanghai Jiao Tong University,
	Minhang 200240, China (e-mail: search4meaning@sjtu.edu.cn; yongpeng.wu@sjtu.edu.cn; zhaiguangtao@sjtu.edu.cn; ningliu@sjtu.edu.cn; zhangwenjun@sjtu.edu.cn)(Corresponding author: Yongpeng Wu.).}
}

\author{\IEEEauthorblockN{Jie Feng, Yongpeng Wu, Guangtao Zhai, Ning Liu, and Wenjun Zhang}
}
\maketitle

\begin{abstract}
Virtual reality (VR) is making waves around the world recently. However, traditional video streaming is not suitable for VR video because of the huge size and view switch requirements of VR videos. Since the view of each user is limited, it is unnecessary to send the whole 360-degree scene at high quality which can be a heavy burden for the transmission system. Assuming filed-of-view (FoV) of each user can be predicted with high probability, we can divide the video screen into partitions and send those partitions which will appear in FoV at high quality. Hence, we propose an novel strategy for VR video streaming. First, we define a quality-of-experience metric to measure the viewing experience of users and define a channel model to reflect the fluctuation of the wireless channel. Next, we formulate the optimization problem and find its feasible solution by convex optimization. In order to improve bandwidth efficiency, we also add adaptive modulation to this part. Finally, we compare our algorithm with other VR streaming algorithm in the simulation. It turns out that our algorithm outperforms other algorithms.
\end{abstract}


\section{Introduction}
\vspace{1.5 ex}
\subsection{Background}
Virtual reality (VR) promises to revolutionize the way people interact with media. Providing users with an immersive experience in a virtual world by transmitting panoramic video streams, VR is drawing more and more attentions and has been applied to many fields such as entertainment, medical application, education, and so on. A research report on Chinese VR market indicates that the VR market is expected to exceed \$15.9 billion in 2019 and reach \$30 billion in 2020 [1].

However, current technologies for streaming try to fetch all the portion of the chunks in the same quality, which results in a high bandwidth utilization [2] and a limitation on content generation. It is clear that traditional video streaming strategies cannot be directly applied to VR video steaming. This is because the traditional method does not consider VR's unique features such as its huge size and view switch requirements. Moreover, video quality, transmission errors and delay should be considered especially when a VR video is transmitted over unstable band-limited wireless channels. Hence, transmission algorithms are required to save resources and ensure the quality of experience (QoE) of the users.

\begin{figure}[htbp]
\centerline{\includegraphics{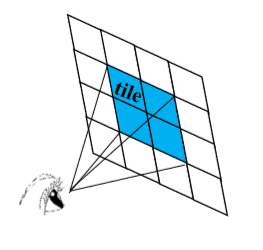}}
\caption{Illustration of the tile division.(Blue tiles represents the FoV area.)}
\label{fig}
\end{figure}

\vspace{1.5 ex}
\subsection{Related work}
The most common way adopted by researches on VR video streaming is to divide the initial large imagine into smaller pieces (often called tiles). Given that each time users can only watch a certain part of a panoramic video which is called Field of View (FoV), there is no need to transmit all the tiles in high quality. Fig.1 shows an illustration of tile division. And lots of bandwidth can be saved by only setting tiles in FoV to high quality. Many works have been done on this basis. Hosseini proposed an adaptive view-aware bandwidth-efficient VR video streaming framework based on the tiling features of spatial relationship description [3]. A rate adaption algorithm was proposed in [4]. Its basic idea is to fetch the invisible portion of a video at the lowest quality based on users¡¯ head movement prediction, and to adaptively decide the video playback quality for the visible portion based on bandwidth prediction. Liu et al. considered unequal error protection for FoV and formulated the inherent error resilient VR video transmission problem into a joint source and channel coding problem [5].

Besides, there are also many researches on resource allocation with multiple users. Chen et al. proposed a novel VR model based on multi-attribute utility theory and formulate the resource allocation problem as a noncooperative game. A distributed algorithm based on the machine learning framework of echo state networks was proposed to solve the resource allocation problem [6]. Since keeping sending real time tracking information about users¡¯ FoV will lead to a significant burden for a communication system, some work has been done on FoV prediction. Zhu et al. established a model to predict the saliency maps and scanpaths for the users'  head movement and eye motion data and scan paths data, respectively [7].

\vspace{1.5 ex}
\subsection{Our work}
The intention of our work is to present a bandwidth-efficient VR video transmission strategy with relatively high QoE under changeable channel conditions. In particular, we propose our model and design an algorithm that can adaptively adjust encoded bitrate according to the wireless channel condition.

We first define a QoE metric. Two issues are considered when defining the QoE metric. One is the video quality, and the other is the stall time during adjacent chunks due to bad channel condition. Intuitively, high encoded bitrate can lead to severe stall time. In our QoE metric, we adopt the idea of saliency map to reflect the video quality and use the stall time definition in [4]. The QoE metric shows a trade-off between video quality and play-back delay.

Secondly, we formulate the optimization problem to maximize the QoE metric. With some relaxations, we convert the original problem to a convex one. Based on the solution of the convex problem, we use our algorithm to get the feasible solution of the original problem.

Next, we provide simulations to examine the proposed algorithm. Numerical results indicate that the proposed algorithm performs better comparing to two existing algorithms in [4].  Moreover, adaptive modulation brings additional performance gains.

In general,the contributions of our work are summarized as follows:
\begin{itemize}
\item We formulate the QoE metric for VR video streaming on the basis of saliency map.
\item We convert the optimization problem of QoE to a relaxed convex one and propose an algorithm to find the feasible solution.
\item We employ adaptive modulation to take the fading effect of wireless channel into consideration when designing the transmission strategy.
\item Numerical results are provided to illustrate the effectiveness of the proposed algorithm. The proposed algorithm achieves significant performance gains comparing to the existing algorithms in [4].
\end{itemize}

The rest of this paper is organized as follows. In Section \uppercase\expandafter{\romannumeral2}, we propose our system model. In Section \uppercase\expandafter{\romannumeral3}, we formulate the QoE maximization problem and provide an algorithm to find its feasible solution. Numerical results are provided in Section \uppercase\expandafter{\romannumeral4}. Finally, conclusions are drawn in Section \uppercase\expandafter{\romannumeral5}.

\emph{Notation:} Lowercase letters $x$ and uppercase letters $X$ denote scalars. The Euclidean norm operator is denoted by $\left \| \cdot  \right \|$. Let $z\sim \mathcal{CN}(0,\sigma ^2)$ denote a complex Gaussian random variable $z$ with zero mean and variance $\sigma ^2$.

\vspace{1.5 ex}
\section{System Model}
In this section, we introduce the system model considered in this paper. In general, we incorporate saliency map, wireless channel, and adaptive modulation scheme to the system model in [4].

\vspace{1.5 ex}
\subsection{Evaluation metrics for video quality}
We assume the video is composed of $K$ chunks. Each chunk lasts $L$ seconds. Then, the total video duration is $KL$. For each chunk $k\in \left \{ 1,2,...,K \right \}$, there are $N$ tiles in total. The tile $i$ within chunk $k$ can be encoded at the rate $R_{i,k}\in \left \{ R_1,R_2,...,R_{max} \right \}$ where $R_1<R_2<...<R_{max}$. (Note that tiles can only be encoded with optional discrete rates.) Then the total size of the chunk $k$ can be calculated as $L\sum_{1}^{N}R_{i,k}$. The set of the serial number of $M$ tiles in the FoV area of chunk $k$ is denoted as $V_k$. So the size of the FoV area in chunk $k$ is calculated as $L\sum_{i\in{V_k}}R_{i,k}$.

Next, we present evaluation metrics to measure users' viewing experience according to the encoding rate. Note that here we just take encoding rate as a reflection of the picture quality of a VR video without considering the stall time. Functions like $EM\left ( R \right )=aR^b(a>0,0<b<1)$ can be good choices. The constraint on the range of $a$ and $b$ ensures that the function is a strictly increasing convex function, and its first derivative decreases with the increase of independent variable $R$, which is very close to the actual situation. With the encoding rate increasing, users will get higher quality of experience. But this upward trend keeps weakening when the encoding rate reaches higher levels. In terms of human brain characteristics, the perception of an entire scene is achieved in the brain by merging together individual images captured at various fixation points throughout successive saccadic eye movements. In general, tiles with distinctive features are more likely to attract users and play a greater role in brain imaging. Therefore, instead of regarding $\underset{i\in V_k}{min}R_{i,k}$ as a dominant optimization metric as in [4],  we introduce the concept of saliency map in this paper. In real experiments [8], users' eye-gaze data can be stored and analyzed to build a saliency map in witch tile $i$ in chunk $k$ is attached with weight $w_{i,k}$. Finally, the evaluation metrics for video quality of chunk $k$ can be written as:
\begin{equation}
\sum_{i\in V_k}w_{i,k}EM\left ( R_{i,k} \right ).
\end{equation}

\vspace{1.5 ex}
\subsection{Wireless channel model}
To meet the increasing demand of VR, we need to consider the wireless transmission of VR in a large space. Hence, we should consider the impact of the wireless channel on transmission. In our model, we consider a channel with additive white Gaussian noise affected by both large scale fading and small scale fading
\begin{equation}
h\sim \alpha h_r,
\end{equation}
$\alpha$ is subject to a uniform distribution, representing the large scale fading. $h_r\sim \mathcal{CN}(0,1)$ represents the small scale fading.

\vspace{1.5 ex}
\subsection{Adaptive modulation}
Adaptive modulation is widely used in today's wireless transmission and has been adopted in our transmission strategy to improve bandwidth utilization. The basic idea is the most bandwidth-efficient modulation is used at a guaranteed low bit error ratio (BER). Fig.2 shows BER for various modulations as a function of short-term average SNR at the receiver. Suppose the symbol rate and transmitted power are fixed, we can determine the modulation mode according to the detected channel condition. Let $m_k$ be the spectral efficiency for chunk $k$, and it can be mathematically expressed as:
\begin{equation}
m_k=max\left \{ i:S(P_e,i)\leqslant \frac{\left \| h \right \|^2P_t}{BN_o} \right \}.
\end{equation}
In the above equation, $i\in\left \{ 1,2,4,6 \right \}$ represents the spectral efficiency of BPSK, QPSK, 16-QAM and 64-QAM. $S(P_e,i)$ gives the threshold value of SNR for modulation mode corresponding to $i$ with BER equal to $P_e$. $B$ is the channel bandwidth.

\begin{figure}
\centerline{
\includegraphics[width=3.5in]{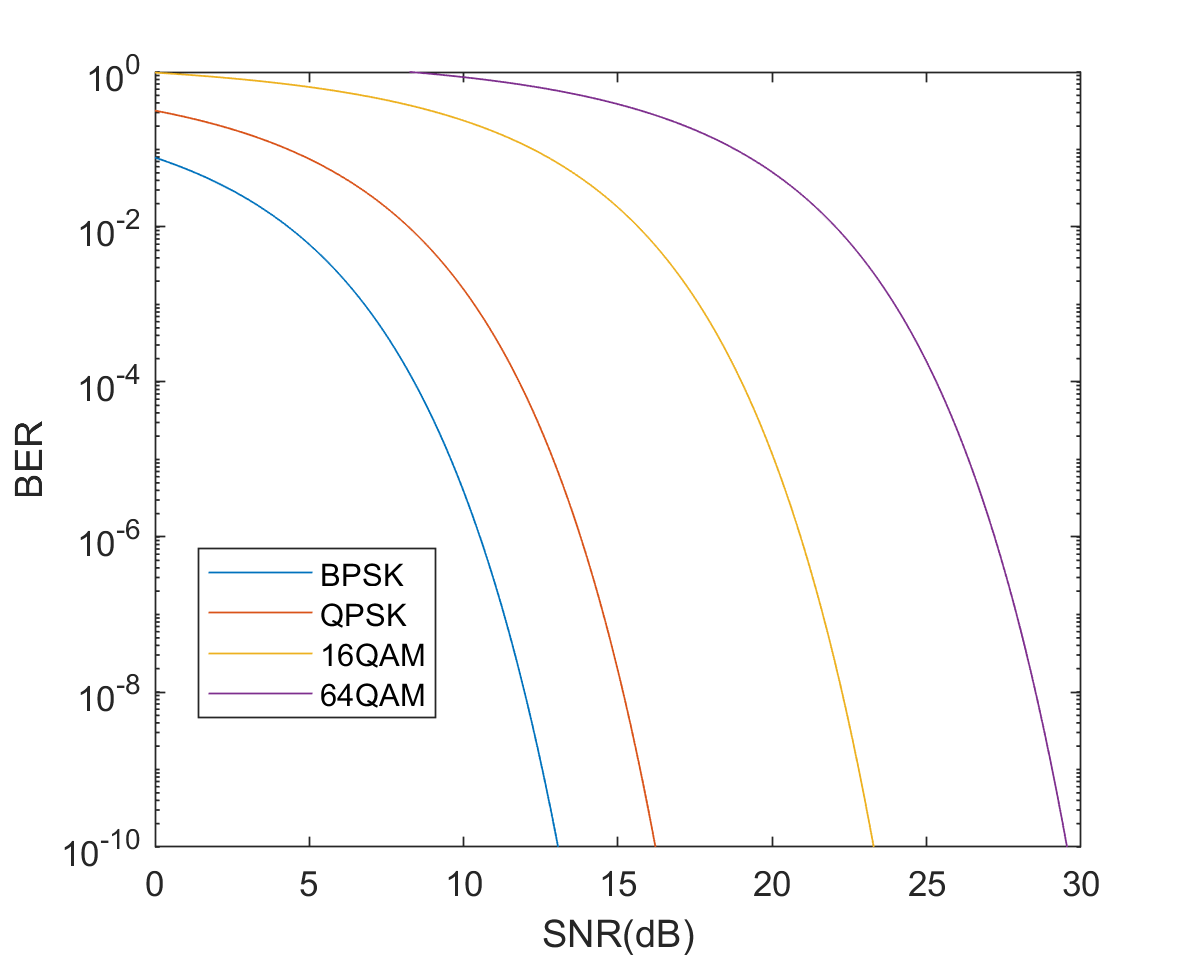}}
\caption{BER for various modulations level as a function of short-term average SNR.}
\label{fig}
\end{figure}

\vspace{1.5 ex}
\subsection{Stall time}
VR video chunks are first downloaded to users' local buffer before they are played. In practice, the buffer capacity is limited. However, considering the large amount of VR video data, the downloading speed is too limited for the cache to hold multiple chunks at the same time. So we do not consider the limitation of buffer capacity on our model. Chunk $k$ can only be downloaded after chunk $k-1$ has been downloaded. Let $t_k$ denotes the start time of downloading chunk $k$ and suppose there is no gap between downloading two chunks. Then, for $k>1$, we have
\begin{equation}
t_k=t_{k-1}+\frac{L\sum_{i=1}^{N}R_{i,k-1}}{m_{k-1}B}=t_1+L\sum_{j=1}^{k-1}\frac{\sum_{i=1}^{N}R_{i,j}}{m_jB},
\end{equation}
where the bandwidth $B$ also equals to the symbol rate. $m_{k-1}B$ is the downloading rate. $L\sum_{i=1}^{N}R_{i,k-1}$ is the total size of chunk $k$.

Chunk $k$ can only be played after it is downloaded to the buffer and chunk $k-1$ has been played. Let $\widetilde{t}_k$ denote the play time of chunk $k$. For $k>1$, we have
\begin{equation}
\begin{aligned}
\widetilde{t}_k&=max\left \{ \widetilde{t}_{k-1}+L,t_k+\frac{L\sum_{i=1}^{N}R_{i,k}}{m_{k}B}\right \}
\\&=max\left \{ \widetilde{t}_{k-1}+L, t_1+L\sum_{j=1}^{k}\frac{\sum_{i=1}^{N}R_{i,j}}{m_jB}\right \}.
\end{aligned}
\end{equation}
Since the start time is considered to be when users see the first chunk being played, we simply set $t_1=\widetilde{t}_1=t_2=0$.

Stall time will appear if the play of chunk $k-1$ ends before chunk $k$ is downloaded to the buffer. Denote the stall time between chunk $k-1$ and chunk $k$ as $\Delta t_k$ ($\Delta t_1=0$) and we have

\begin{equation}
\begin{aligned}
\Delta t_k &= \widetilde{t}_k-\widetilde{t}_{k-1}-L
\\&=max\Big \{ 0, L\sum_{j=1}^{k-1}\frac{\sum_{i=1}^{N}R_{i,j}}{m_jB}-\widetilde{t}_{k-1}-L\Big \}
\\&=max\Big \{\!0\!,\! L\!\sum_{j\!=\!1}^{k\!-\!1}\!\frac{\sum_{i\!=\!1}^{N}R_{i,j}}{m_jB}\!-\!(\!\widetilde{t}_{k\!-\!2}+\!L\!+\!\Delta t_{k\!-\!1}\!)\!-\!L\!\Big \}
\\&=max\Big\{ 0, L(\sum_{j=2}^{k-1}(\frac{\sum_{i=1}^{N}R_{i,j}}{m_jB}-\frac{\Delta t_j}{L}-1)\\&\quad+\frac{\sum_{i=1}^{N}R_{i,k}}{m_kB}-1 )\Big\}.
\end{aligned}
\end{equation}

\vspace{1.5 ex}
\section{QoE Optimization}
\vspace{1.5 ex}
In this section, we first define our QoE metrics and formulate the optimization problem. Next, we design an algorithm to find the feasible solution.

\vspace{1.5 ex}
\subsection{QoE maximization problem}
The quality of users' experience should be determined in two ways. One is the encoding rate at FoV area which directly affects the quality of the video scene, and the other is the stall time when video is played. Obviously, we need a steaming strategy that makes the former term larger while the latter smaller. Thus, we define the QoE metrics of chunk $k$ as follow:
\begin{equation}
QoE=\sum_{i\in V_k}w_{i,k}EM\left ( R_{i,k} \right )-\lambda\Delta t_k.
\end{equation}
\\
At the same time, we would like to make the following explanations to this optimization target:
\begin{itemize}
\item $\lambda>0$ is the weight of stall time $\Delta t_k$. As $\lambda$ increases, the stall time and average bitrate in FOV will both decrease in the optimal solution.
\item In our model, our optimization target is specific to each chunk. And the optimal result for chunk $k$ is based on chunk $1,\cdots ,k-1$. So the final solution may not be the optimal for the whole video. However, due to technical limitations, it is impossible to achieve channel detection and FoV prediction for a long time like a whole video duration. Hence, it makes sense to narrow down the optimization time interval.
\item When solving the optimization problem, we suppose that the FoV area is correctly predicted. However, it is quite possible to predict a wrong area. And this is also the reason why we encode tiles outside FoV at bitrate $R_1$ (lowest optional rate).\\

\end{itemize}

So far, we formulate our optimization problem for chunk $k$
\begin{align*}
\underset{R_{i,k},i\in V_k}{\mathbf{maximize}}& \qquad\qquad \sum_{i\in V_k}w_{i,k}EM\left ( R_{i,k} \right )-\lambda\Delta t_k\\
\mathbf{s.t.}& \quad \Delta t=max\Big\{ 0, L(\sum_{j=2}^{k-1}(\frac{\sum_{i=1}^{N}R_{i,j}}{m_jB}-\frac{\Delta t_j}{L}-1)\\&\qquad\quad\;+\frac{\sum_{i=1}^{N}R_{i,k}}{m_kB}-1 )\Big\},\\
&\quad R_{i,k}\in \left \{ R_1,\cdots ,R_m \right \}.
\end{align*}
in which constrain (6) is also equivalent to the following two inequalities
\begin{equation}
\Delta t_k \geq 0,
\end{equation}

\begin{equation}
\Delta t_k \geq L(\sum_{j=2}^{k-1}(\frac{\sum_{i=1}^{N}R_{i,j}}{m_jB}\!-\!\frac{\Delta t_j}{L}\!-\!1)\!+\!\frac{\sum_{i=1}^{N}R_{i,k}}{m_kB}\!-\!1 ).
\end{equation}

And the reformulated optimization problem is
\begin{align*}
\underset{R_{i,k},i\in V_k}{\mathbf{maximize}}& \qquad\;\; \sum_{i\in V_k}w_{i,k}EM\left ( R_{i,k} \right )-\lambda\Delta t_k\\
\mathbf{s.t.}& \quad \Delta t_k \geq 0,\\
&\quad\Delta t_k \geq L(\sum_{j=2}^{k-1}(\frac{\sum_{i=1}^{N}R_{i,j}}{m_jB}\!-\!\frac{\Delta t_j}{L}\!-\!1)\!\\&\qquad\qquad+\!\frac{\sum_{i=1}^{N}R_{i,k}}{m_kB}\!-\!1 ),\\
&\quad R_{i,k}\in \left \{ R_1,\cdots ,R_m \right \}.
\end{align*}

\vspace{1.5 ex}
\subsection{Feasible solution}
The optimization problem mentioned above is not a typical convex one because the range of variables $R_{i,k}$ is discrete. In this subsection, we present our algorithm that finds a feasible solution to the problem.

Suppose now the range of $R_{i,k}$ is continuous from $R_1$ to $R_m$, the original problem reduces to a convex optimization problem where the optimal value is denoted as $R_{i,k}^\ast$. If we choose $R_{i,k}=max\left \{ R_j:R_j\leq R_{i,k}^\ast \right \}$, we will get a solution which is close to the optimal solution. However in most cases, this will still be a waste of bandwidth. Thus, we provide an algorithm with higher bandwidth efficiency and solutions closer to the optimal.

\begin{algorithm}[t]
\caption{The main algorithm for finding feasible solutions for chunk $k$ to the QoE opimization problem.}
    \begin{algorithmic}[1]
    \State \textbf{Initialization:} $T_k=0.$
    \For{each $i\in V_k$}
      \State $R_{i,k}\leftarrow max\left \{ R_j:R_j\leq R_{i,k}^\ast \right \};$
      \State $\overline{R}_{i,k}\leftarrow min\left \{ R_j:R_j\geq R_{i,k}^\ast \right \};$
      \State $T_k\leftarrow T_k+(R_{i,k}^\ast-R_{i,k});$
    \EndFor

    \If{$T_k = 0$}
        \State \textbf{exit};
    \EndIf

    \For{$j=1, \cdots, M$}
      \State $i\leftarrow \left \{ q\!:\!w_{q\!,k}\; \rm{is\,the}\, j^{th}\,\rm{ largest\, weight\, for\, tiles\, in}\, V_k \right \};$
      \State $T_k\leftarrow T_k-(\overline{R}_{i,k}-R_{i,k});$
      \If{$T_k\geq 0$}
        \State $R_{i,k}\leftarrow \overline{R}_{i,k};$
      \Else
        \State \textbf{break};
      \EndIf
    \EndFor
    \end{algorithmic}
\end{algorithm}

The basic idea of the algorithm is that we try to increase the encoding rate according to the priority of weights in the saliency map without increasing the delay. Since $EM(\cdot)$ is convex, the improvement in QoE slows down as the encoding rate increases. And generally the overall quality improvement at FOV is better than only one highlighted. Thus, rather than setting the encoding rate of one tile two levels up, we prefer to set both two tiles one level up.

\begin{figure*}
\begin{minipage}{0.5\textwidth}
  \centerline{\includegraphics[width=1\textwidth]{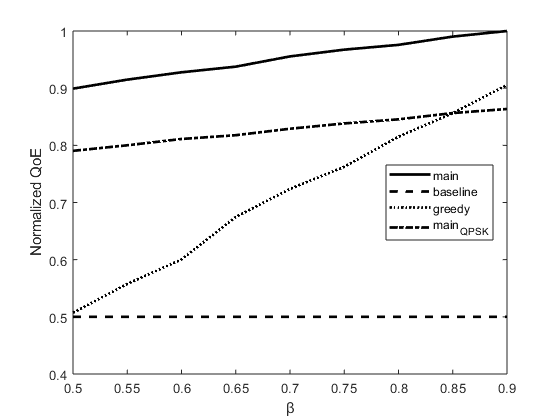}}
  \caption{The variation of the normalized QoE with $\beta$.}
\end{minipage}
\hfill
\begin{minipage}{0.5\textwidth}
  \centerline{\includegraphics[width=1\textwidth]{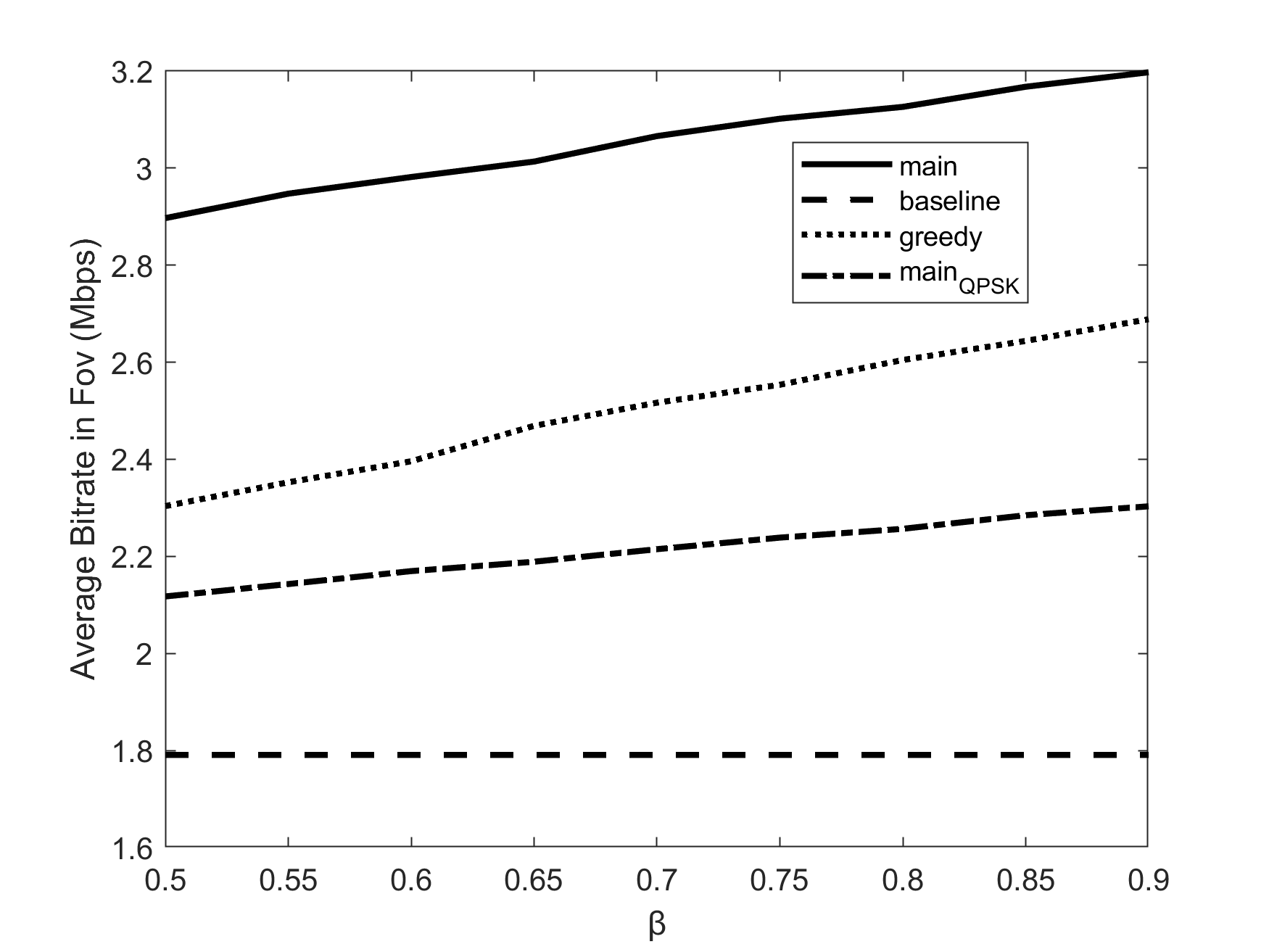}}
  \caption{The variation of the average bitrate in FoV with $\beta$.}
\end{minipage}
\vfill
\begin{minipage}{0.5\textwidth}
  \centerline{\includegraphics[width=1\textwidth]{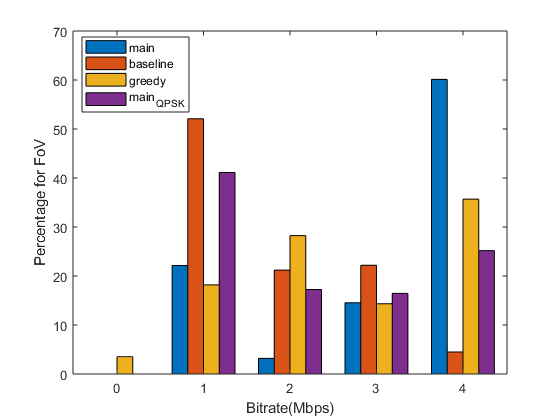}}
  \caption{The distribution of the bitrates for FoV when $\beta=0.8$.}
\end{minipage}
\hfill
\begin{minipage}{0.5\textwidth}
  \centerline{\includegraphics[width=1\textwidth]{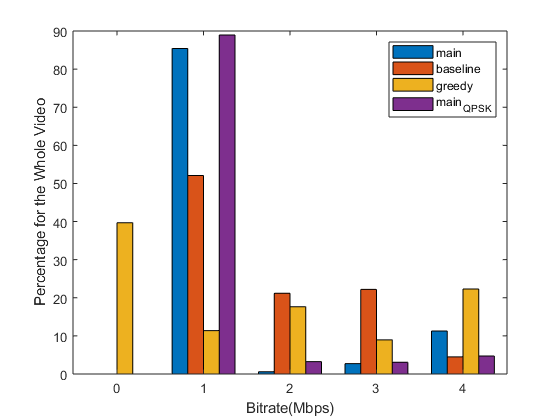}}
  \caption{The distribution of the bitrates for the whole video when $\beta=0.8$.}
\end{minipage}
\label{fig}
\end{figure*}
\vspace{1.5 ex}
\section{Simulations and Results}
In this section, we evaluate our algorithm in simulations, and compare it with other VR streaming algorithms. (Note that the curves obtained by our algorithm is marked as ``main" in the figure notes.)

\vspace{1.5 ex}
\subsection{Parameter setup}
In simulation, the VR video is divided into $4\times 8$ tiles. And the FoV area is consists of $2\times 3$ tiles. The duration of each chunk is 2 seconds. And for the number of chunks $K=1000$, the whole length of the VR video is around 36 minutes. And we set each tile's rate $R_{i,j}\in \left \{1, 2, 3, 4  \right \}$Mbps. Half the weights $w_{i,k}$ are set at 1 and the rest are randomly set between 1 and 2.

The channel bandwidth is 20Mbps with a 25\% fluctuation. We change $\alpha$ every 40 seconds and $h_r$ every 2 seconds based on their distribution to simulate the channel condition.

The probability of FoV being correctly predicted is denoted as $\beta$. And in our simulations, we set the value of $\beta$ to vary between 0.5 and 0.9.

We also simulate algorithms mentioned in [4] for VR streaming. In the baseline algorithm [9], all the tiles within a chunk is encoded at a same rate although most tiles will not be seen. And greedy algorithm [10] only sends those tiles with highest probability to be the apart of FoV and ignores the rest ones, which may result in users seeing blank tiles when FoV prediction fails. Apart from these, we also use our algorithm on the situation when bits are only QPSK modulated [4] to see how adaptive modulation will benefit the system. Unless otherwise noted, by saying ``our algorithm'', we refer to our algorithm with adaptive modulation.

\vspace{1.5 ex}
\subsection{Results and discussions}

We observe from Fig. 3 that as the FoV prediction probability $\beta$ increases, both our algorithm and greedy algorithm increase QoE. The performance of our algorithm is better than greedy algorithm and baseline algorithm. As $\beta$ increases, greedy algorithm approaches our algorithm. This is because greedy algorithm selects a larger potential FoV area than our algorithm at the cost of rest tiles unsent. In other words, in greedy algorithm, bandwidth resources are centralized for a certain portion of tiles rather than all of them. When $\beta$ is low, blank tiles may be sent in FoV by using greedy algorithm, resulting in low QoE. However, as $\beta$ increases and FoV prediction error decreases, greedy algorithm shows a faster growth. The curve of baseline algorithm remains flat because its performance only depends on the channel condition. Comparing the performance of our algorithm under different modulation schemes, we verify that adaptive modulation is superior to QPSK. Additionally, the advantage of adaptive modulation over QPSK keeps expanding as $\beta$ increases.

The average bitrate of tiles in the FoV area vs. $\beta$ is shown in Fig. 4. It is illustrated in Fig. 4 that except for baseline algorithm, the average bitrate increases monotonically with $\beta$ for other algorithms. And our algorithm outperforms other algorithms by at least 23\%. Compared with QPSK modulation, our algorithm with adaptive modulation provides significant improvement of average bitrate in FoV.

Fig. 5 shows the distribution of tiles with different encoding levels in the FoV area. Of all the algorithms, our algorithm provides the highest proportion of 4Mbps tiles. For baseline algorithm, 1Mbps tiles dominate. One unique feature of greedy algorithm is that it is the only algorithm that have 0Mbps tiles. This is because blank tiles will appear in the actual FoV area when prediction error occurs according to the encoding scheme of greedy algorithm. It is also illustrated in Fig. 5 that with adaptive modulation, our algorithm fetches more percentage of 4Mbps tiles and fewer 1 or 2Mbps tiles than QPSK modulation.

Fig. 6 plots the percentage of tiles encoded in different levels for the whole video. For algorithms that encode all the tiles, 1Mbps tiles have the largest proportion. While in greedy algorithm, the dominant tiles are blank tiles. The percentage of 4Mbps tiles of greedy algorithm exceeds that of our algorithm because it selects a larger potential FoV area than our algorithm at the cost of rest tiles unsent. However, greedy algorithm still performs poorer than our algorithm in QoE because FoV can not be predicted 100\% correctly and blank tiles will cause severe impact on QoE.

\vspace{1.5 ex}
\section{Conclusion}
In this paper, we propose a novel strategy for VR video streaming based on adaptive modulation. We define a QoE metric according to features of brain imaging and provide an algorithm to find feasible solution of the QoE maximization problem. We evaluate our algorithm in simulations to compare it with baseline algorithm and greedy algorithm. In addition, we use our algorithm under QPSK modulation to see the advantage of adaptive modulation. Numerical results indicate that our algorithm outperforms other algorithms. Besides, compared to QPSK modulation, our algorithm improves the QoE by at least 13\% using adaptive modulation. In the future, we expect to evaluate our algorithm in real VR videos.

\end{document}